# Non-invasive single-shot 3D imaging through a scattering layer using speckle interferometry

ATUL S. SOMKUWAR,[1] BHARGAB DAS,[2] VINU R.V,[1] YONGKEUN PARK,[3] RAKESH KUMAR SINGH[1,*]

[1] Applied and Adaptive Optics Laboratory, Department of Physics, Indian Institute of Space Science and Technology (IIST), Valiamala, Trivanrdum, 695547, Kerala , India
[2] Advance Materials and Sensors Division, CSIR-Central Scientific Instruments Organisation, Chandigarh, 160030, India
[3] Department of Physics, Korea Advanced Institute of Science and Technology, Daejeon, 305-701, Republic of Korea
*Corresponding author: krakeshsingh@iist.ac.in

***Optical imaging through complex scattering media is one of the major technical challenges with important applications in many research fields, ranging from biomedical imaging, astronomical telescopy, and spatially multiplex optical communications. Although various approaches for imaging though turbid layer have been recently proposed, they had been limited to two-dimensional imaging. Here we propose and experimentally demonstrate an approach for three-dimensional single-shot imaging of objects hidden behind an opaque scattering layer.*** **We demonstrate that under suitable conditions, it is possible to perform the 3D imaging to reconstruct the complex amplitude of objects situated at different depths.**

## 1. INTRODUCTION

When an object is situated behind a scattering layer, both bright-field incoherent imaging and conventional coherent imaging are not suited to extract useful information about the object from the scattered wave. The inhomogeneity of refractive indexes in the scattering layer diffuses the field information of the object into a highly disordered complex speckle pattern [1].This was long thought of as a randomizing process which limits the penetration depth of optical imaging techniques [2]. Intensive research efforts have been made to address this fundamental, yet practical, problem and many different techniques have been put forward. Significant breakthroughs have been made using methods exploiting deterministic linear processes of light transport through turbid media [3-5]; imaging objects through complex media has been realized [6-21].However, most of these techniques are limited by specific requirements such as presence of a guiding star or known object [8, 22], nonlinear materials or nonlinear processes [14, 15,23], unscattered ballistic light [10-12], long acquisition or processing sequences [24,25], ultrasonic encoding [26, 27], and exhaustive searching algorithm [8, 9].Techniques for microscopic imaging and quantitative phase contrast mapping of objects in turbid microfluidic channels by using the concepts of digital holography were also explored [28-30]. In particular, flow of medium in microfluidic channel provides a solution and exploited to acquire set of holograms with uncorrelated speckle patterns to be incoherently averaged.

Recently, non-invasive imaging through a scattering layer has been performed using speckle autocorrelation and phase retrieval methods [31]. More recently, an improved approach based on speckle autocorrelation was proposed [32], wherein the object information is reconstructed non-destructively by utilizing the ergodic property of speckle patterns in diffusive regime. This circumvents the requirement of lengthy angular scanning procedures [31] and hence reduces the measurement time significantly. Unfortunately, the aforementioned techniques do not provide three-dimensional (3D) imaging capability, as explicitly pointed out in Ref. [32].Thus the actual position of the object behind the scattering medium cannot be ascertained. Measurement of the actual depth of an object hidden behind a scattering medium is crucial in diverse research areas and applications such as biomedical imaging, deep tissue microscopy, colloidal imaging, astronomy etc. This limitation of depth reconstruction for the techniques based on the speckle memory effect is mainly owing to the inability to measure the full phase space [33]. The full $\{x, k\}$ phase space measurement is implemented with help of two independent CCD recordings, windowed Fourier transforms through an SLM, and phase retrieval algorithm. However, this previous idea proposed by Takahashi and Fleischer [33] uses a complex experimental arrangement with multiple CCD recordings, windowed Fourier transforms, and phase retrieval algorithm.

In a recent communication, we proposed and experimentally validated a rather simple concept of recovering complex coherence function of the random field from two-point intensity correlation measurement [34]. The complex coherence function in the far field is connected with random source structure by a Fourier transform relation. Therefore, this relationship can be exploited to develop a new imaging technique through random scattering medium which has wide range of applications covering laboratory environment to astronomical imaging. Novelty of this technique lies with its ability to exploit the statistical feature of the laser speckle for direct 3D imaging purpose without resorting to any wavefront correction and iterative process. Here, we celebrate the randomness of the laser speckle rather than removing it. Moreover, the proposed work is capable to detect complex 3D field of the object from a single shot measurement of the far field laser speckle. To the best of our knowledge, we are the first to propose and demonstrate the imaging of 3D complex objects with full depth information through a scattering medium by using a single CCD recording.

Hence, this communication reports a novel and simple non-invasive single-shot imaging method that allows the 3D reconstruction of objects behind a scattering medium. In order to retain the depth reconstruction capability, the imaging technique must retrieve the complete object field information (both amplitude and phase) at the plane of scattering media. Thus, we propose and experimentally demonstrate a coherent 3D imaging method based on the concepts of speckle holography [35] together with two-point intensity correlation (or speckle correlation). The use of holographic principles for imaging through random media dates back to the invention of lasers [36-38]. Similarly, two-point intensity correlation or fourth-order correlation has been used in stellar interferometry for reconstruction of an image of a distant star. In recent years, it has also been used to recover 3D objects [39], and detection of gratings hidden by a diffuser [40].Here, we demonstrate that the combination of the principles of holographic imaging and two-point intensity correlation can provide a foundation for 3D imaging through turbid media. To illustrate the potential of our method, we experimentally validate the reconstruction of two sets of objects separated by two different distances of 10 mm and 15 mm with full amplitude and phase information. In another separate experiment, we also demonstrated depth imaging of a real world object located at a distance of z=155mm from the random scattering plane.

## 2. RESULTS

### A. Principle

A schematic of the proposed method is shown in Fig. 1(a), wherein object information, scrambled by scattering medium, is Fourier transformed by a lens and combined with a reference speckle field $U_R(r_2)$. In the absence of a reference speckle pattern, the intensity distribution $I(r_2)$ recorded by the CCD detector allows us to estimate the fourth order correlation function as

$$C(r_2, r_2 + \Delta r_2) = \langle \Delta I(r_2) \Delta I(r_2 + \Delta r_2) \rangle, \quad (1)$$

where $\langle \cdot \rangle$ represents ensemble average, $I(r_2) \equiv U(r_2)U^*(r_2); U(r_2)$ being the object speckle filed at the CCD plane. Further, $\Delta I(r_2) = I(r_2) - \langle I(r_2) \rangle$ is the spatial fluctuation of the recorded intensity with respect to the average value, $r_2$ is the transverse spatial coordinate in the CCD plane. In general, the ensemble averaging is performed over multiple recorded speckle patterns. However, assuming a spatially stationary and ergodic speckle pattern, the temporal averaging can be replaced by spatial averaging with high fidelity. This also enables single-shot imaging capability. Furthermore, assuming that the speckle field from the scattering medium follows a Gaussian statistics, the fourth-order correlation of equation (1) can be expressed in terms of a complex coherence function or second-order correlation function $W(\Delta r_2)$ as [1, 41,42].

$$C(\Delta r_2) \propto |W(\Delta r_2)|^2, \quad (2)$$

where $W(\Delta r_2) \equiv \langle U(r_2)U^*(r_2 + \Delta r_2) \rangle_s$ and $\langle \cdot \rangle_s$ represents the spatial averaging.

Thus, under suitable conditions, the intensity correlation function of the recorded speckle pattern is analogous to the squared modulus of the complex coherence function. According to the van Cittert-Zernike theorem, the complex coherence function $W(\Delta r_2)$ can be used to reconstruct the object information since it is equivalent to the Fourier transform of the source structure at the scattering medium [41,42].However, equation (2) measures only the squared modulus of the complex coherence function, which is not enough for the reconstruction of the object information. Importantly, the lost phase information of $W(\Delta r_2)$ is need to be recovered. The use of iterative phase-retrieval algorithms was suggested by Bertolotti *et al.*[31]and Katz *et al.*[32] for this purpose. Although promising, iterative phase retrieval algorithms can be computationally demanding with large number of iterations, and they can sometimes stagnate before converging to an acceptable solution and/or exhibit a relatively slow convergence rate. In addition, iterative techniques are not yet adapted/or highly computationally involved for the reconstruction of complex fields.

In this communication, we used the basic approach of interferometry in order do the same task. As shown in Fig. 1(a), a reference speckle pattern is introduced by illuminating an independent ground glass with an off-axis point source. In the presence of the reference speckle pattern, the resultant intensity correlation function, under the conditions of egodicity and Gaussian statistics, can be expressed as

$$\left|\widetilde{W}(\Delta r_2)\right|^2 = |W(\Delta r_2)|^2 + |W_a(\Delta r_2)|^2 + W(\Delta r_2)W_a^*(\Delta r_2) + W^*(\Delta r_2)W_a(\Delta r_2), \quad (3)$$

where $\widetilde{W}(\Delta r_2)$ and $W_a(\Delta r_2)$ are the complex coherence functions of the resultant speckle and reference speckle fields, respectively. Equation (3) reveal that both amplitude and phase information of object coherence function is now retained. Since the reference speckle field (i.e. $W_a(\Delta r_2)$) is generated by an off-axis point source, the complex coherence function $W(\Delta r_2)$ due to the object can be easily separated from other redundant factors in equation (3)by Fourier domain filtering approach (Supplementary information) and hence allows us to reconstruct the optical information hidden by the scattering medium. Furthermore, if we assume that the optical field at the scattering medium is an interference pattern due to an object and a reference beam as shown in Fig. 1(b), the above proposed technique reconstructs this interference pattern with high fidelity. This will be clearer when we present our experimental results. Since the actual object information is encoded in an interference pattern, the wavefront due to object can be reconstructed from this interference, which is further utilized in retrieving the depth resolved images. The complex amplitude of the object at different depths is obtained by performing a beam propagation operation based on the angular spectrum method [43].

### B. Experimental Validation

In order to experimentally validate our proposed method, we first utilized two sets of objects separated by a depth and also separated in the transverse plane as shown in Figs. 2(a) and (b). The longitudinal distance between the letter 'O' and 'W' is $\Delta z = 10\ mm$ and between the shapes 'Star' and 'heart' is $\Delta z = 15\ mm$. Fourier holograms of both the set of objects are numerically generated as described in the supplementary information and are shown in Figs. 2(c) and (d). These Fourier holograms retain the complete 3D object information. All the objects used in the experimental study are of dimension ~ 0.25 mm × 0.15 mm. Figure 1(c) shows the actual experimental setup used for the measurements. Linearly polarized light from a He-Ne laser ($\lambda$=632.8 nm, Melles Griot, 25-LHP-925) was used as a coherent light source. The beam, after spatial filtering and collimation is then split by a beam splitter $BS_1$ into two beams to form a Mach-Zehnder interferometer (MZI). In one arm of the interferometer, we have a spatial light modulator (SLM, Holoeye LC-R 720, Reflective Type; Pixel pitch: 20 μm) which displays the numerically generated Fourier holograms with the help of polarizing beam splitter PBS and half-wave plate HWP.

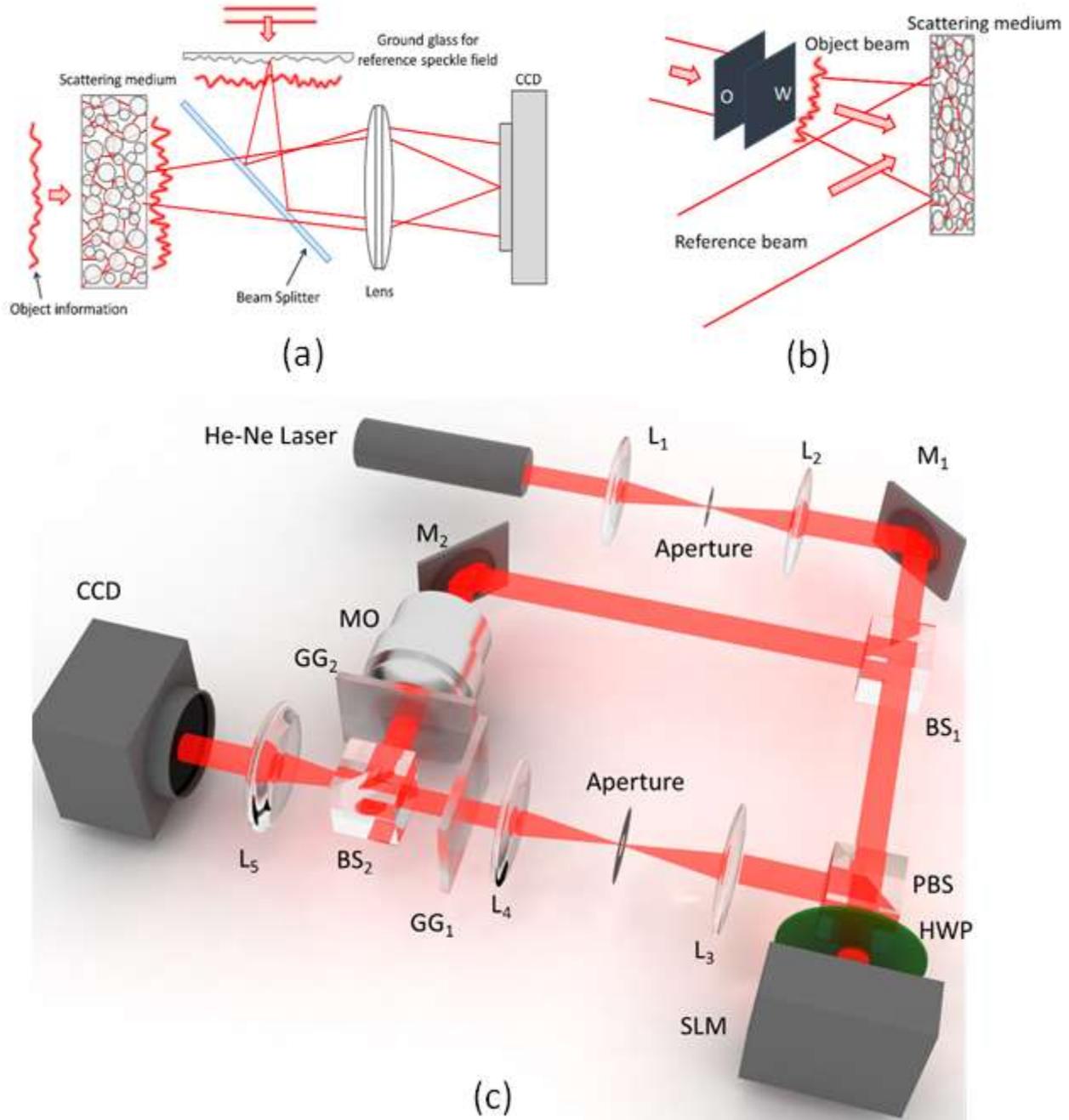

Fig 1. (a) Schematic of the proposed optical method used for the reconstruction of object situated behind a scattering medium with full depth sectioning capability; (b) The optical field at the scattering medium is considered to be an interference pattern due to an object and a reference beam; (c) Schematic of the actual experimental set-up

The hologram is then imaged on to a scattering medium with the help of lenses $L_3$ and $L_4$. Ground glasses (DG20/120/MD, Thorlabs) with a thickness of 3 mm are considered as the scattering medium which scrambles the Fourier hologram information into a highly distorted speckle pattern. The ground glasses are held static during the recording process. The scattered light from the ground glass $GG_1$ is then Fourier transformed using the lens $L_5$. Our main objective is to retrieve the Fourier hologram information at the plane of the scattering medium from a single CCD recording by using the principle described in the previous section. In order to retrieve this Fourier hologram information at the plane of the ground glass $GG_1$, a reference speckle pattern is introduced using a second ground glass $GG_2$ in the other arm of the MZI. The reference speckle field is synthesized by illuminating the ground glass $GG_2$ with an off-axis point source produced by a microscope objective MO, which is then Fourier transformed by the lens $L_5$. The resultant speckle hologram due to the two beams is then recorded by a high-resolution monochrome CCD (Prosillica GX2750, 2750×2200 pixels and 4.54 μm pixel pitch). Once the Fourier holograms are faithfully recovered from the single CCD speckle image, the 3D object information can be reconstructed through dc-suppression and beam propagation techniques.

Figures 2(e) and (f) show the raw intensity images recorded by the CCD for the objects shown in Figs. 2(a) and (b), respectively.

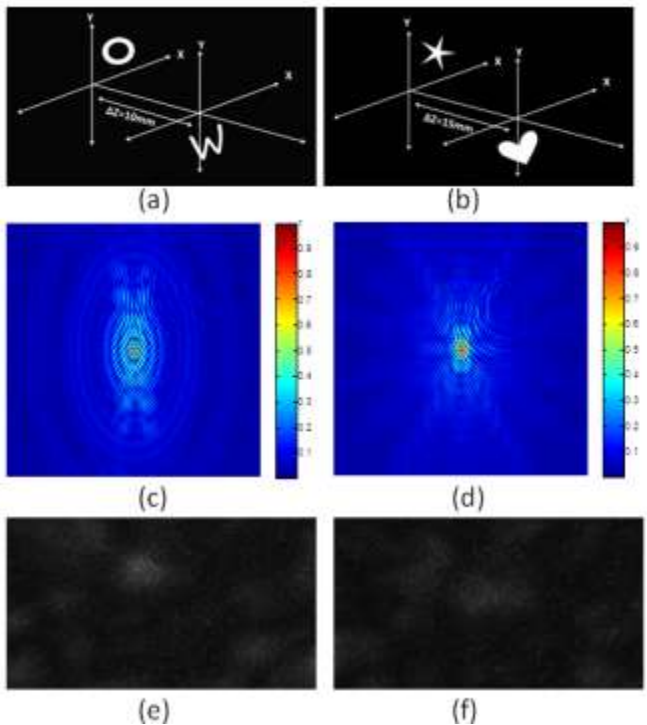

Fig. 2. Two sets of objects for experimental validation (a) Letters 'O' and 'W' separated by a longitudinal distance of $\Delta z = 10\ mm$, (b) Shapes 'Star' and 'Heart' by $\Delta z = 15\ mm$. (c) & (d) are the numerically generated Fourier holograms of (a) & (b) to be displayed in the SLM. (e) & (f) are raw recorded CCD images for the Fourier holograms shown in (c) and (d) respectively. The color bars show normalized intensity.

The object information in the form of Fourier hologram is scrambled by the scattering medium and does not provide any visual information. The recorded intensity images only display low-contrast random patterns. Spatially averaged two-point intensity correlation functions are then calculated from the recorded intensity images which are equivalent to $\left|\widetilde{W}(\Delta r_2)\right|^2$ in Equation (3). The complex coherence function $W(\Delta r_2)$ of the optical field at the scattering medium is encoded within this two-point intensity correlation function. We now invoke the van Cittert-Zernike theorem to reconstruct the optical field at the plane of the scattering medium which is equivalent to the Fourier holograms of the objects. This is obtained by performing the Fourier transform of the calculated intensity correlations. Because reference coherence is a complex function with a uniform amplitude profile and a linear phase factor as explained by the van Cittert-Zernike theorem, the reconstructed Fourier hologram is situated at an off-axis position, and requires a spatial filtering and centring operation. The retrieved Fourier holograms are then utilized to reconstruct the two objects situated at different depths. Figures 3(a)-(j) show the depth resolved amplitude and phase information of the two set of objects. The multiple in-focus images appearing in Figs. 3(a) and (b) are due to the inherent property of Fourier hologram which also reconstructs the conjugate object information. Numerical beam propagation technique based on the angular spectrum method is used for retrieving the complex amplitudes at different depths as shown in Figs. 3(c) and (f). The reconstruction of the objects with full amplitude and phase information confirms the fact that the proposed method can faithfully image through an opaque scattering medium. This is further validated by reconstructing the depth resolved object information from the Fourier holograms without the scattering medium. The experimental results are shown in Figs. 4(a)-(j), wherein the Fourier hologram from the SLM is directly imaged onto the CCD without the intervening scattering medium. The reconstruction results of Figs. 4 show high degree of similarity with the results retrieved in presence of the scattering medium as shown in Figs. 3. This consolidates the potential of our single-shot imaging technique through an opaque medium.

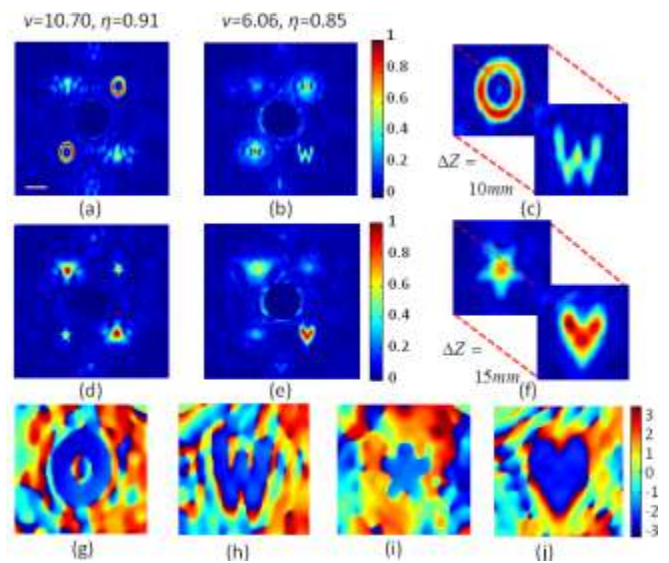

Fig. 3: Reconstructed amplitude and phase information through a scattering medium; (a) & (b) are the amplitude information of the two letters, (c) 3D representative diagram showing focusing of the two objects with depth separation of Δz =10 mm; Similarly, (d)-(f) are the reconstructed amplitude information for the shapes 'star' and 'heart'. (g)-(j) are the reconstructed phase information for the two sets of objects. Color bars in (a), (b), (d) & (e) represent the normalized amplitude and in (g)-(j) represent phase in radians. The scale bar shown in (a) is of size 0.2 mm and it is same for (b), (d) & (e). *Visibility (v)* and *Reconstruction Efficiency (η)* values are shown just above the figures.

The reconstruction quality is further assessed by considering two important parameters: *visibility (v)* and *reconstruction efficiency (η)*. The visibility of the target reconstruction is defined as the extent to which the reconstruction is distinguishable from background noise. It is measured as the ratio of the average image intensity level in the region corresponding to the transparent area (*i. e.* signal region) to the average background intensity level. The signal region is identified by using global threshold technique. The calculated *visibility* values for the images shown in Figs. 3(a) and 4(a) are 10.70 and 11.26, which corresponds to the reconstructed images in the presence and without the scattering medium, respectively. Similarly, the *v* values of the reconstructed objects in Figs. 3(b) and 4(b) are 6.06 and 6.1 respectively. Likewise, the *reconstruction efficiency (η)* can be defined as the ratio of measured power in the signal region of the image to the sum of the measured powers in signal and background regions. The $\eta$ values for the same two sets of reconstructed images (Figs. 3(a) &4(a) and 3(b) & 4(b)) are found out to be (0.91 and 0.92) and (0.85 and 0.86) respectively. These *visibility* and *reconstruction efficiency* results show very good correlation between the images retrieved with and without the scattering medium demonstrating the feasibility of the proposed method.

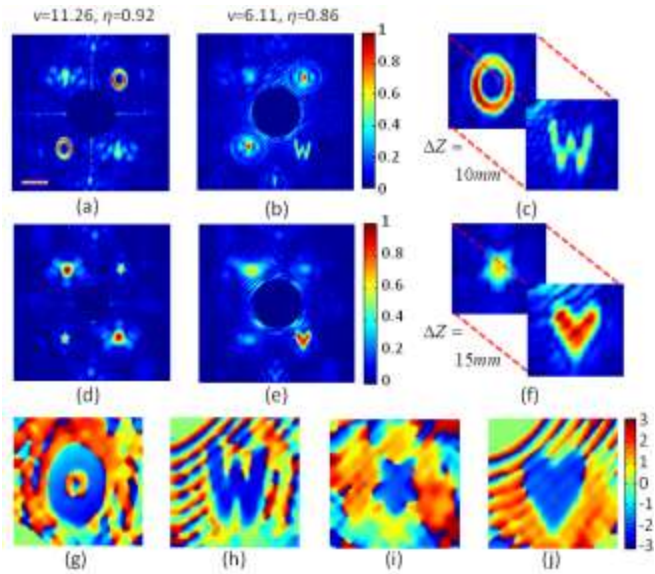

Fig. 4: Reconstructed amplitude and phase information without using a scattering medium; (a) & (b) are the amplitude information of the two letters, (c) 3D representative diagram showing focusing of the two objects with depth separation of Δz =10 mm; Similarly, (d)-(f) are the reconstructed amplitude information for the shapes 'star' and 'heart' for Δz =15 mm. (g)-(j) are the reconstructed phase information for the two sets of objects. Color bars in (a), (b), (d) & (e) represent the normalized amplitude and in (g)-(j) represent phase in radians. The scale bar shown in (a) is of size 0.2 mm and it is same for (b), (d) & (e). *Visibility (v)* and *Reconstruction Efficiency (η)* values are shown just above the figures.

Furthermore, in order to demonstrate the applicability of our presented technique for 3D imaging through scattering media of real world object, we have modified the experimental geometry described in Fig. 1(c). The SLM and the 4*f* geometry used to image the Fourier hologram at the scatterer plane is replaced with a second Mach-Zehnder interferometer which generates the hologram of an actual object that illuminates the ground glass $GG_1$ (details described in Supplementary Information).The object used for this experimental study is the letter "V" of dimension 5mm×4mm printed on a transparency and placed at a distance of 155mm from scatterer. The random intensity pattern from $GG_1$ is superposed with the reference random pattern from $GG_2$ and Fourier transformed using lens $L_5$. Fig. 5(a) shows the raw intensity image recorded by the CCD. The hologram of the object at the scatterer is recovered from the recorded random intensity pattern of Fig. 5(a) using two point intensity correlation and by calculating the Fourier transform of correlation function as described in the previous case. The complex amplitude of the hidden object at different depths from the scattering medium is then retrieved by using angular spectrum based beam propagation technique. Figs. 5(b) and 5(c) show reconstructed amplitude and phase information, respectively, at different depths. The front figure in each case shows the retrieved information at the correct depth of z = 155 mm. The visibility *(v)* and reconstruction efficiency *(η)* for the retrieved image at the actual depth in presence of scattering medium are 6.02 and 0.86 respectively. For comparison, the *v* and *η* values without the scattering medium are 6.68 and 0.87 respectively. As before, we obtained very good parallelism of the reconstruction results. The experimental demonstration of real world object of Fig. 5 shows the applicability of the proposed method for 3D imaging of objects hidden behind the scattering medium.

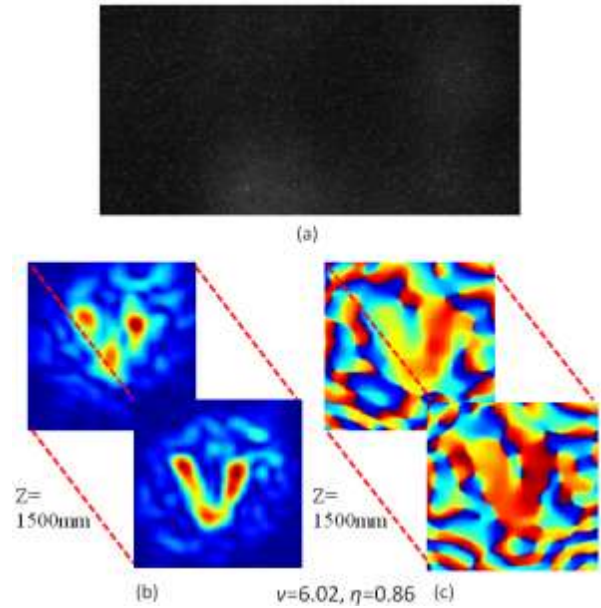

Fig. 5: Reconstructed amplitude and phase information through a scattering medium; (a) raw recorded image for an off axis hologram, (b) reconstructed amplitude information of the object at a distance 155mm from scattering medium and its defocused information at z= 1500mm from object plane; (c) reconstructed phase information of the object at a distance 155mmfrom scattering medium and its defocused information at z=1500mm from object plane. *Visibility (v)* and *Reconstruction Efficiency (η)* values at the actual depth are shown just below the figure.

## 3. DISCUSSION

We presented a speckle holography technique for 3Dimaging of objects completely hidden behind an opaque scattering layer. A single-shot recording of an intensity image is sufficient to recover high quality object information. The performance of the present technique depends on the requirement of a coherent reference speckle pattern; a simple and fast reconstruction algorithm can be employed if the experimental optical geometry guarantees a coherent reference speckle pattern. Importantly, 3D imaging through a turbid layer was demonstrated because the optical information at the scattering layer is assumed to be an interference pattern. Thus, the proposed technique can also be adapted to a situation where the optical information just before the scattering layer can be considered as an in-line hologram [44], and will have the similar advantages.

Furthermore, it should be noted that the quality of the reconstructed image is determined by a number of important factors. Firstly, the spatial resolution of the retrieved image is strongly dictated by the finite correlation length of the scattering medium used in the experiments. Ideally it is assumed that speckle patterns produced by the scattering medium have a sharply peaked autocorrelation function. When this assumption is true, the resolution of the recovered object depends on the size of the Fourier hologram and $f$ number of the Fourier transformation lens. Furthermore, the accuracy of the recovered object is also dependent on the validity of spatial stationarity and Gaussian statistics assumptions of the speckle field. Full object information recovery may not be achieved when the spatial stationarity is ensured only in patches in the CCD plane. Another important factor that governs the reconstruction quality of the presented technique is that the speckles in the recording plane should be well resolved by the CCD in order to meet the sampling criterion. This means the speckle grain size should be greater than two CCD pixels. This condition is satisfied in our experimental setup by placing an aperture of suitable dimension at the plane of the scattering medium. Thus the actual size of the hologram contributing to the speckle formation is limited by this aperture. In order to appreciate the novelty of our work, it is important to differentiate our idea from the work presented in Ref.[38] wherein the idea of imaging the hologram obscured by the diffuser is applied. This is carried out by using impulse response function of diffraction limited imaging system under illumination of delta correlated diffused light. In order to compensate the effect of speckle noise due to limited size of the imaging system and its influence on the imaging of hologram, time average over the speckle field intensity is performed by rotating the diffuser.

In summary, we have experimentally demonstrated a single-shot 3D imaging technique through a highly scattering medium employing the concepts of speckle holography and two point intensity correlation. The demonstrated technique has the unique ability to retrieve the complex filed behind a scattering medium resulting into the reconstruction of actual position of the object. Thus our method provides a robust and practical solution to implement the 3D complex field imaging through an opaque scattering medium for wide range of applications.

**Acknowledgment**. Part of this work is carried out under the IIST supported fast track project.